\documentclass{lhep}       

\journal{LHEP}
\vol{xx}
\jyear{2018}
\pages{xxx} 

\received{xx January 2018}
\published{xx March 2018}

\def\be{\begin{equation}}
\def\ee{\end{equation}}
\def\bea{\begin{eqnarray}}
\def\eea{\end{eqnarray}}

\usepackage{graphics} 
\usepackage{amsmath}
\usepackage{amssymb}
\usepackage{times}
\usepackage{dsfont}
\usepackage{cancel}
\usepackage{graphicx}
\usepackage{pictexwd,dcpic}
\usepackage{epigraph}
\usepackage{mathrsfs}
\usepackage{epstopdf}
\newtheorem{thm}{Theorem}

\begin{document}

\title{Scalar models of formally interacting non-standard quantum fields in Minkowski space-time}

\author{Andreas Aste\auno{1}\auno{2}}
\address{$^1$Department of Physics, University of Basel, Klingelbergstrasse 82, 4056 Basel, Switzerland}
\address{$^2$Paul Scherrer Institute, Forschungsstrasse 111, 5232 Villigen PSI, Switzerland}

\begin{abstract}
For decades, a lot of work has been devoted to the problem of constructing a non-trivial
quantum field theory in four-dimensional space-time.
This letter addresses the attempts to construct an algebraic quantum field theory
in the framework of non-standard theories like hyperfunction or ultra-hyperfunction
quantum field theory. For this purpose model theories of formally
interacting neutral scalar fields are constructed and some of their characteristic properties
like two-point functions are discussed. The formal self-couplings are obtained
from local normally-ordered analytic redefinitions of the free scalar
quantum field, mimicking a non-trivial structure of the resulting Lagrangians and
equations of motion.
\end{abstract}

\maketitle


\section{Introduction}
Standard relativistic quantum field theory (QFT) in the sense of G\r arding and Wightman \cite{PCT} uses
the Schwartz space $\mathcal{S} (\mathds{R}^4)$ of rapidly decreasing $C^\infty$-functions
as a test function space, and in this context
a quantum field $\mathcal{O}$ is an operator-valued distribution, expressing the
fact that $\mathcal{O} (f)$ is an (unbounded) operator defined on a dense subset $\mathcal{D}$
of a Hilbert space $\mathcal{H}$ for all $f \in \mathcal{S} (\mathds{R}^4)$. The underlying symmetry of the
theory  is the Poincar\'e group $\mathcal{P}^\uparrow_+$, i.e. the semidirect product of the abelian group
of time-space translations $T_{1,3}$ and the restricted Lorentz group $SO^+(1,3)$, or, to be more precise, 
the covering group $\bar{\mathcal{P}}^\uparrow_+ = T_{1,3} \rtimes SL(2,\mathds{C})$
when fermionic fields are included \cite{PCT}.\\

The free neutral scalar field $\varphi(x)$ with the Wightman two-point function
$\langle 0 | \varphi(x) \varphi (y) | 0 \rangle = i \Delta^+ (x-y)$ given by the positive-frequency
Pauli-Jordan $C$-number distribution $\Delta^+$ which has the Fourier transform
\begin{equation}
\hat{\Delta}^+ (k) = \int \frac{d^4 x}{( 2 \pi)^2} \, \Delta^+ (x) e^{ikx} =
- \frac{i} {2 \pi} \Theta (k^0) \delta (k^2 - m^2) \, , \label{twopointf}
\end{equation}
where $kx=k_ \mu x^\mu = k^0 x^0 - \vec{k} \cdot \vec{x} =
k_0 x^0+k_1 x^1 + k_2 x^2 + k_3 x^3=k^0 x^0-k^1 x^1 - k^2 x^2 -k^3 x^3$
and $k^0=(\vec{k}^{\, 2} + m^2)^{1/2}$,
provides a simple example for a quantum field associated with a free particle of mass $m$
in $3+1$ space-time dimensions. As operator-valued distributions, all $\varphi(f)$ act on a common
dense set of the standard bosonic Fock-Hilbert space $\mathcal{F}$ with a non-degenerate vacuum
represented by a normalized state vector $| 0 \rangle$, as discussed in many textbooks.

Using Schwartz functions as test function space, it is possible to express the causal structure of
QFT by the help of (anti-)commutation relations for (fermionic) bosonic operators smeared with test
functions having compact support. E.g., a neutral scalar field fulfills the commutation relation \cite{Scharf1,Scharf2}
\begin{equation}
[ \varphi(f), \varphi (h) ] = 0
\end{equation}
if the compact supports of the test functions $f,h \in \mathcal{S} (\mathds{R}^4)$ are spacelike to each other,
i.e. if all $x \in \mbox{supp} (f)$ and $y \in \mbox{supp} (h)$ are spacelike separated: $(x-y)^2 < 0$.

Distribution theory is a linear theory and no associative product of two distributions extending the product
of a distribution by a smooth function can be defined. In the case of the free field operator $\varphi$,
a partial solution of the problem is offered by the normal ordering of field operators which corresponds to a
recursive point-splitting regularization described in a formal manner as follows
\begin{equation}
:\varphi(x): = \varphi(x) \,  ,
\end{equation}
\begin{equation}
:\varphi(x)^2: =\lim_{y \rightarrow x} [ \varphi(x) \varphi(y) - \langle 0 | \varphi(x) \varphi(y) | 0 \rangle ] \,  ,
\end{equation}
\begin{displaymath}
:\varphi(x)^n: =\lim_{y \rightarrow x} [: \varphi(x)^{n-1}: \varphi(y)
\end{displaymath}
\begin{equation}
- (n-1) \langle 0 | \varphi(x) \varphi(y) | 0 \rangle :\varphi(x)^{n-2}: ] \,  .
\end{equation}

\noindent The normally ordered product $:\varphi(x)^n:$ is an operator-valued distribution
again \cite{Constantinescu,Strocchi},
as well as the tensor product $:\varphi(x)^n:$$:\varphi(y)^n:$.
Accordingly, also Wick polynomials defined as finite sums of normally ordered products
\begin{equation}
p(x) = \sum \limits_{n=0}^N a_n \frac{: \varphi (x)^n :}{n!}
\end{equation}
are densely defined operator valued distributions in the Fock-Hilbert space $\mathcal{F}$
with well-defined correlation distributions \`a la $\langle 0 | p(x) p(y) | 0 \rangle$.

But a problem arises from the fact the two-point function (\ref{twopointf}) shows a singular behaviour
on the light cone. In $s \ge 2$ space dimensions, $\langle 0 | \varphi(x) \varphi (y) | 0 \rangle$
has a local singularity of the form $[(x-y)^2]^{(1-s)/2}$, and therefore extending Wick polynomials to
infinite power series
\begin{equation}
\rho (x) =  \sum \limits_{n=0}^\infty a_n \frac{: \varphi (x)^n :}{n!} \label{Wickpower}
\end{equation}
and calculating correlation distributions $\langle 0 | \rho (x) \rho (y) | 0 \rangle$ yields an essential
singularity at the origin, and consequently objects like $\rho (x)$ will not be tempered.

Despite this problem, Nagamachi and Mugibayashi \cite{NagaMugi1,NagaMugi2} were able to show that the concept of localization
can be implemented in the non-standard framework of hyperfunction quantum field theory (HFQFT) without making
use of compactly supported test functions, but in terms of {\emph{Fourier hyperfunctions}.
The space of Fourier hyperfunctions is the dual of the space of rapidly decreasing holomorphic functions.
One of the characteristics of this space is that it is topologically invariant under Fourier transformations
as is the case for the spaces $\mathcal{S} (\mathds{R}^n)$, but it does {\emph{not}} contain test functions of
compact support.
But introducing smaller test function spaces than $\mathcal{S}$ seems to be desirable in view of the fact
that in four-dimensional space-time no non-trivial standard
quantum fields have ever been constructed, indicating that the axioms of QFT based on tempered distributions
are too narrow.

It is the aim of this paper to give some insight into the problems which arise when interactions
are taken into account in relativistic quantum field theory by studying some specific examples
of formally interacting field theories, leaving aside the mathematical technicalities involved in the
theory of Fourier hyperfunctions and performing formal calculations.

In this context, a main result of Nagamachi and Br\"uning \cite{NagBru} shall be quoted
here as a theorem:
\begin{thm}
Let $\varphi$ be a free massive neutral scalar field and $\{a_n \}_{n \in \mathds{N}_0}$ a
sequence of real numbers satisfying the condition
$\lim_{n \rightarrow \infty} [|a_n|^2 / n!]^{1/n} = 0$.
Then the Wick power series
\begin{displaymath}
\rho (x) =  \sum \limits_{n=0}^\infty a_n \frac{: \varphi (x)^n :}{n!}
\end{displaymath}
is a hyperfunction quantum field,
but not a standard quantum field if infinitely many of the coefficients $a_n$ are non-zero.
\label{theorem1}
\end{thm}
Of course the condition  $\lim_{n \rightarrow \infty} [|a_n|^2 / n!]^{1/n} = 0$ ensures
that for $z \in \mathds{C}$
\begin{equation}
h(z) =\sum \limits_{n=0}^\infty a_n^2 \frac{z^n }{n!}
\end{equation}
is an entire function.

The hyperfunction approach to quantum field theory has been extended
to ultra-hyperfunction approaches with even more restricted test function spaces during
the last years indeed \cite{Vindas}. But also these approaches do not seem to lead anywhere
from a physical point of view. 


\section{Example of a non-standard quantum field}
Before tackling models of formally interacting quantum fields in Minkowski space, we briefly discuss a typical
example of a hyperfunction quantum field $\Phi$ given by the Wick power series containing a free neutral
field $\varphi$ \cite{Schroer,Aste} 
\begin{equation}
\Phi (x) = : e^{\lambda \varphi (x)} : = \sum \limits_{n=0}^\infty \frac{\lambda^n}{n!} :\varphi (x)^n :
\end{equation}
with some length parameter $\lambda$.
The corresponding two-point function is
\begin{displaymath}
w^{\Phi} (x-y) = \langle 0 | \Phi (x) \Phi (y) | 0 \rangle
\end{displaymath}
\begin{displaymath}
= \langle 0 |  \sum \limits_{n=0}^\infty \frac{\lambda^n}{n!} :\varphi (x)^n :
 \sum \limits_{m=0}^\infty \frac{\lambda^m}{m!} :\varphi (y)^m : | 0 \rangle
\end{displaymath}
\begin{equation}
=   \sum \limits_{n=0}^\infty \frac{i^n \lambda^{2n}}{n!} \bigl[ \Delta^+ (x-y) \bigr]^n
= e^{i \lambda^2 \Delta^+ (x-y)} \, ,
\end{equation}
since for combinatorial reasons
\begin{equation}
\langle 0 | : \varphi (x)^n : : \varphi (y)^m : | 0 \rangle = \delta_{nm} n! \bigl[ i \Delta^+ (x-y) \bigr]^n \, .
\end{equation}
Considering the massless case, the
positive-frequency Pauli-Jordan distribution is given in configuration space by
\begin{equation}
\Delta^+_0(x) = \frac{i}{4 \pi^2} \frac{1}{(x_0 - i0)^2 -\vec{x}^2} \, ,
\end{equation}
hence the massless two-point function becomes
\begin{equation}
w^{\Phi}_0 (x-y) 
= e^{i \lambda^2 \Delta^+_0 (x-y)}
=  \displaystyle \exp \biggl( {-\frac{\lambda^2}{4 \pi^2 (x^2 - ix_0 0)}} \biggr) \, .
\end{equation}
Due to its essentially singular behaviour, $w^\Phi_0$ (and the massive $w^\Phi$) cannot be a tempered
distribution in $\mathcal{S}' (\mathds{R}^4)$. Tempered distributions can always be represented as a
finite sum of distributional derivatives of continuous functions of polynomial growth.

Trying to evaluate the two-point function of an expression like, e.g.,
\begin{equation}
\Psi (x) = : 2 \ln (1+ \lambda \varphi (x)/2) : = \sum \limits_{n=1}^\infty \frac{(-1)^{n+1}
\lambda^n}{2^{n-1} n} : \varphi (x)^n :
\end{equation}
formally leads to
\begin{equation}
\langle 0 | \Psi (x) \Psi (y) | 0 \rangle = \sum \limits_{n=1}^\infty \frac{\lambda^{2n} n!}{2^{2n-2} n^2}
 \bigl[ i \Delta^+ (x-y) \bigr]^n \, ,
\end{equation}
an expression which does not converge in any sense.


\section{Formal interactions through point transformations of the classical Lagrangian of the free neutral scalar field}
The following exercises on models of formally interacting fields will shed some additional  light on some of the
comments in the introduction. They may also serve as interesting examples for point transformations in lectures
on Lagrangian field theory. 
\subsection{Massive free and formally interacting field: classical and quantum aspects}
The Lagrangian density $\mathcal{L}_0$ of the non-interacting classical real scalar field $\varphi$
\begin{equation}
\mathcal{L}_0 (\varphi, \partial_\mu \varphi) = \frac{1}{2} \partial_\mu \varphi \partial^\mu \varphi - \frac{m^2}{2} \varphi^2
\end{equation}
can be cast in a less familiar form by a local point transformation
with a real parameter $\lambda$ 
\begin{equation}
\varphi(x) = \lambda^{-1} \tan \bigl( \lambda \psi (x) \bigr) \, .
\end{equation}
A point transformation
\begin{equation}
\mathcal{L}_1 (\psi, \partial_\mu \psi) = \mathcal{L}_0 (\varphi (\psi), \partial_\mu \varphi (\psi))
\end{equation}
leaves the form of the Euler-Lagrange equations
\begin{equation}
\partial_\mu \frac{\mathcal{L}_0}{\partial \partial_\mu \varphi} -
\frac{\partial{\mathcal{L}_0}}{\partial \varphi} = 0
= \partial_\mu \frac{\mathcal{L}_1}{\partial \partial_\mu \psi} -
\frac{\partial{\mathcal{L}_1 }}{\partial \psi}
\end{equation}
invariant. The free field $\varphi$ obeys the Klein-Gordon equation
\begin{equation}
\partial_\mu \frac{\partial \mathcal{L}_0}{\partial \partial_\mu \varphi} -
\frac{\partial{\mathcal{L}_0}}{{\partial \varphi}}
= 0 = \Box \varphi + m^2 \varphi \, .
\end{equation}
With
\begin{equation}
\partial_\mu \varphi (x)  = \frac{\partial_\mu \psi (x)}{\cos^2 \bigl( \lambda \psi (x) \big) }
\end{equation}
follows
\begin{equation}
\mathcal{L}_1 (\psi, \partial_\mu \psi) = \frac{\partial_\mu \psi \partial^\mu \psi}{2 \cos^4 ( \lambda \psi)}
-\frac{m^2}{2 \lambda^2} \tan^2 (\lambda \psi) \, . \label{Lagra1}
\end{equation}
The Euler-Lagrange equations for $\psi$ are obtained from
\begin{equation}
\frac{\partial \mathcal{L}_1}{\partial \psi} = - \frac{m^2}{\lambda} \frac{\sin (\lambda \psi)}{\cos^3 ( \lambda \psi) }
\end{equation}
and
\begin{displaymath}
\partial_\mu \frac{\partial \mathcal{L}_1}{\partial \partial_\mu \psi}
= \partial_\mu \Bigl( \frac{\partial^\mu \psi}{\cos^4 (\lambda \psi)} \Bigr)
\end{displaymath}
\begin{equation}
= \frac{\Box \psi}{\cos^4 (\lambda \psi)} + \frac{4 \lambda \sin (\lambda \psi )}{\cos^5 ( \lambda \psi)}
\partial_\mu \psi \partial^\mu \psi \, ,
\end{equation}
i.e. one has
\begin{equation}
\Box \psi = - 4 \lambda \tan ( \lambda \psi ) \partial_\mu \psi \partial^\mu \psi - \frac{m^2}{\lambda} \sin ( \lambda \psi)
\cos ( \lambda \psi) \, . \label{ELcomp}
\end{equation}
Of course, solutions of the seemingly complicated equation of motion (\ref{ELcomp})
can be generated by taking a solution of the real Klein-Gordon equation $( \Box + m^2 ) \varphi = 0$
and calculating $\psi = \lambda^{-1} \arctan ( \lambda \varphi)$. However, the field theories defined
by $\mathcal{L}_{0,1}$ are not completely equivalent, since a solution $\varphi = \lambda^{-1} \tan (\lambda \psi)$
of the Klein-Gordon equation corresponds to a denumerable discrete set of solutions
$\{ \psi + m \pi / \lambda \mid m \in \mathds{Z} \}$. There exists an infinity of parallel $\psi$-worlds.

Regarding the fact that $\varphi \simeq \psi$ for small fields $|\varphi|, |\psi| \ll 1$,
one might be tempted to invoke perturbation theory for the involved Lagrangian density
$\mathcal{L}_1 (\psi, \partial_\mu \psi)$. Expanding
\begin{equation}
\cos^{-4} (\lambda \psi) = 1 + 2 \lambda^2 \psi^2 + o (\psi^4)
\end{equation}
and
\begin{equation}
\tan^2 ( \lambda \psi ) = \lambda^2 \psi^2 + \frac{2}{3} \lambda^4 \psi^4 + o(\psi^6)
\end{equation}
and inserting it in (\ref{Lagra1}) leads to
\begin{displaymath}
\mathcal{L}_1 ( \psi , \partial_\mu \psi) = \frac{1}{2} \partial_\mu \psi \partial^\mu \psi - \frac{m^2}{2} \psi^2
\end{displaymath}
\begin{displaymath}
+ \lambda^2 \psi^2 \partial_\mu \psi \partial^\mu \psi - \frac{m^2 \lambda^2}{3} \psi^4 + \ldots \, .
\end{displaymath}
\begin{equation}
= \mathcal{L}_0 (\psi, \partial_\mu \psi) +  \lambda^2 \psi^2 \partial_\mu \psi \partial^\mu \psi - \frac{m^2 \lambda^2}{3} \psi^4 + \ldots \, .
\end{equation}

Naive power counting indicates that quantization of the Lagrangian density $\mathcal{L}_1$ describing a free
field leads to a non-renormalizable perturbation expansion, since already the lowest quadrilinear
interaction term
\begin{equation}
\mathcal{L}_{1,int}^{(4)} = \lambda^2 \psi^2 \partial_\mu \psi \partial^\mu \psi - \frac{m^2 \lambda^2}{3} \psi^4
\end{equation}
contains the dimension-6 operator $\psi^2 \partial_\mu \psi \partial^\mu \psi$.

Expressing the quantized field $\psi$ in terms of the free quantized field $\varphi$ according to
\begin{equation}
\psi(x) = \lambda^{-1} : \arctan \bigl( \lambda \varphi(x) \bigr): =
\sum \limits_{n=1}^\infty \frac{(-\lambda^2)^{n-1} : \varphi(x)^{2n-1} :}{2n-1} \label{seriespsi}
\end{equation}
as a formal solution of the wave equation (\ref{ELcomp}) does not work.
$\psi$ does not belong to the class of fields according to theorem (\ref{theorem1}),
and it is impossible to calculate a corresponding meaningful two-point function using
expression (\ref{seriespsi}).


\subsection{Formally interacting massless model}
In this section, $\varphi$ represents the (quantized) free {\emph{massless}} neutral scalar field
(with the two-point function $\langle 0 | \varphi (x) \varphi (y) | 0 \rangle =  i \Delta^+_0 (x-y)$)
fulfilling the distributional wave equation
$\Box \varphi (x) = 0$ following from the (classical) Lagrangian density
\begin{equation}
\tilde{\mathcal{L}}_0 (\varphi, \partial_\mu \varphi) = \frac{1}{2} \partial_\mu \varphi \partial^\mu \varphi \, .
\end{equation}
We introduce a new field $\psi$ which is related to $\varphi$ by the one-to-one correspondence
\begin{equation}
\psi = \lambda^{-1} \sinh (\lambda \varphi) \, .
\end{equation}
Using $\cosh^2 (\lambda \varphi) - \sinh^2 (\lambda \varphi) = 1$ and
\begin{equation}
\partial_\mu \psi (x) = \cosh \bigl( \lambda \varphi (x) \bigr) \partial_\mu \varphi (x)
\end{equation}
leads to the classically equivalent Lagrangian density for $\psi$
\begin{displaymath}
\tilde{\mathcal{L}}_1 ( \psi, \partial_\mu \psi) =\tilde{\mathcal{L}}_0 (\varphi, \partial_\mu \varphi)
\end{displaymath}
\begin{equation}
= \frac{1}{2} \frac{\partial_\mu \psi \partial^\mu \psi}{\cosh^2 (\lambda \varphi)}
=   \frac{1}{2} \frac{\partial_\mu \psi \partial^\mu \psi}{1+ \lambda^2 \psi^2} \, .
\end{equation}
In the present case,
\begin{equation}
\lambda^{-1} :\sinh \bigl( \lambda \varphi (x) \bigr) : =  \sum \limits_{n=0}^\infty
\frac{\lambda^{2n}}{(2n+1)!} : \varphi (x)^{2n+1} :
\end{equation}
is a well-defined quantum field with corresponding n-point functions in the sense of HFQFT.


\subsection{Scalar gravity in the absence of matter}
The Lagrangian density for a self-coupled field $h(x)$ in Minkowski space
\begin{equation}
\mathcal{L} = \frac{1}{2} \frac{\partial_\mu h \partial^\mu h}{1 + \lambda h} \label{LagGrav}
\end{equation}
has some interesting properties.
From
\begin{equation}
\partial_\mu \frac{\mathcal{L}}{\partial \partial_\mu h} = \partial_\mu \Biggl( \frac{\partial^\mu h}{1 + \lambda h} \biggl)
= \frac{\Box h}{1 + \lambda h} - \frac{\lambda}{(1+ \lambda h)^2} \partial_\mu h \partial^\mu h
\end{equation}
and
\begin{equation}
\frac{\partial \mathcal{L}}{\partial h} = - \frac{\lambda}{2 ( 1 + \lambda h)^2} \partial_\mu h \partial^\mu h
\end{equation}
follows the equation of motion
\begin{equation}
\Box h = \frac{\lambda}{2} \frac{\partial_\mu h \partial^\mu h}{( 1 + \lambda h)} \, . \label{EOMNord}
\end{equation}
The energy-momentum tensor of the scalar field is given by
\begin{equation}
T^{\mu \nu} = \frac{\partial \mathcal{L}}{\partial \partial_\mu h} \partial^\nu h - g^{\mu \nu} \mathcal{L}
= \frac{\partial^\mu h \partial^\nu h - \frac{1}{2} g^{\mu \nu} \partial_\mu h \partial^\mu h}{1 + \lambda h}
\end{equation}
with the metric tensor $g^{\mu \nu} = diag(1,-1,-1,-1)$.
The trace follows immediately
\begin{equation}
T^\mu_{\mu} = - \frac{\partial_\mu h \partial^\mu h}{1 + \lambda h} \, ,
\end{equation}
i.e. the source of the field $h$ in (\ref{EOMNord}) is proportional to the trace of the energy-momentum tensor of the field
itself. The Lagrangian density (\ref{LagGrav}) defines the ostensibly non-geometrical flat space theory
of Freund and Nambu \cite{FreundNambu}, which has been shown by Deser and Halpern \cite{DeserHalpern}
to be equivalent to the geometrical Nordstr{\o}m theory \cite{Nordstroem}, i.e. the conformally flat metric analog of
Einsteins theory.

Relating $h$ to a new field $\varphi$ by
\begin{equation}
(1 + \lambda \varphi /2) = (1 + \lambda h)^{1/2} \label{hphi}
\end{equation}
results in
\begin{equation}
\partial_\mu \varphi = (1 + \lambda h)^{-1/2}  \partial_\mu h \, ,
\end{equation}
hence
\begin{equation}
\mathcal{L} =  \frac{1}{2} \frac{\partial_\mu h \partial^\mu h}{1 + \lambda h} =
\frac{1}{2} \partial_\mu \varphi \partial^\mu \varphi
\end{equation}
describes a free field $\varphi$ fulfilling the wave equation $\Box \varphi  = 0$.

From the analytic structure of the relation (\ref{hphi}) follows that $h$ cannot be related to the
free massless quantized field $\varphi$ by a Wick power series, despite the classical equivalence of $h$ and $\varphi$.


\section{Conclusions}
The examples of formally interacting scalar field theories presented in this letter illustrate the fact that local non-linear
analytic transformations of a free quantum field, interpreted as an operator valued {\emph{tempered}} distribution
on Schwartz functions, may not result in operator valued tempered distributions, but in a more general kind of distributions
which can be treated in the framework of (ultra-)\-hyper\-func\-tion quantum field theory.

However, even when analytic redefinitions of {\emph{free}} fields are considered only, restrictive analytic conditions must
hold for the field transformations even when they are one-to-one in order for the deformed fields to be interpretable in any
distributional sense. This shows the rigidity of the approach to quantum fields via tempered distributions or (ultra-)hyperfunctions,
and the severe problems of quantum field theory on a classical four-dimensional space-time in general.


\end{document}